\documentclass[10pt,twocolumn,letterpaper]{article}

\usepackage[pagenumbers]{cvpr} % To force page numbers, e.g. for an arXiv version

\usepackage[table,xcdraw,dvipsnames]{xcolor}
\usepackage{times}
\usepackage{epsfig}
\usepackage{graphicx}
\usepackage{amsmath}
\usepackage{amssymb}
\usepackage{url}
\usepackage{multirow}
\usepackage{caption}
\usepackage{subcaption}
\usepackage{xspace}
\usepackage{pifont}% http://ctan.org/pkg/pifont
\usepackage{xparse,xassoccnt}
\usepackage[nocfg]{nomencl}

\usepackage{bm}
\usepackage{multirow}
\usepackage{booktabs}
\usepackage{graphicx}
\usepackage{lscape}
\newcommand{\figref}[1]{Figure~\ref{fig:#1}}
\newcommand{\secref}[1]{Section~\ref{sec:#1}}

\newcommand{\tableref}[1]{Table~\ref{tab:#1}}

\newcommand{\cdms}{\,cd/m$^2$\xspace}

\LetLtxMacro{\originaleqref}{\eqref}
\renewcommand{\eqref}[1]{Eq.~\originaleqref{eq:#1}}

\newcommand{\ind}[1]{\text{#1}}

\newcommand\norm[1]{\left\lVert#1\right\rVert}

\newcommand{\condition}[1]{\emph{#1}}

\definecolor{cvprblue}{rgb}{0.21,0.49,0.74}
\usepackage[pagebackref,breaklinks,colorlinks,citecolor=cvprblue]{hyperref}
\newcommand{\superscript}[1]{\ensuremath{^{\textrm{#1}}}}

\def\paperID{10259} % *** Enter the Paper ID here
\def\confName{CVPR}
\def\confYear{2024}

\title{Training Neural Networks on RAW and HDR Images for Restoration Tasks}

\author{
    Andrew Yanzhe Ke\superscript{2} \hspace{2mm}
    Lei Luo\superscript{1} \hspace{2mm} 
    Xiaoyu Xiang\superscript{1} \hspace{2mm}   
    Yuchen Fan\superscript{1} \hspace{2mm} 
    Rakesh Ranjan\superscript{1} \\
    \vspace{1mm}
    Alexandre Chapiro\superscript{1} \hspace{2mm} 
    Rafa{\l} K. Mantiuk\superscript{2} \\
    \vspace{1mm}
    \superscript{1}Meta \hspace{2mm}
    \superscript{2}University of Cambridge \hspace{2mm}
}

\begin{document}
\maketitle
\begin{abstract}
The vast majority of standard image and video content available online is represented in display-encoded color spaces, in which pixel values are conveniently scaled to a limited range (0--1) and the color distribution is approximately perceptually uniform. In contrast, both camera RAW and high dynamic range (HDR) images are often represented in linear color spaces, in which color values are linearly related to colorimetric quantities of light. While training on commonly available display-encoded images is a well-established practice, there is no consensus on how neural networks should be trained for tasks on RAW and HDR images in linear color spaces. In this work, we test several approaches on three popular image restoration applications: denoising, deblurring, and single-image super-resolution. We examine whether HDR/RAW images need to be display-encoded using popular transfer functions (PQ, PU21, and mu-law), or whether it is better to train in linear color spaces, but use loss functions that correct for perceptual non-uniformity. Our results indicate that neural networks train significantly better on HDR and RAW images represented in display-encoded color spaces, which offer better perceptual uniformity than linear spaces. This small change to the training strategy can bring a very substantial gain in performance, between 2 and 9\,dB. 
\end{abstract}    
\section{Introduction}

Neural networks are typically trained on images represented in \emph{display-encoded} color spaces, such as BT.709 with the sRGB non-linearity \cite{anderson1996proposal}. These are known as \emph{display-encoded} color spaces because they are commonly used to drive displays. HDR images, as well as RAW images containing camera sensor values, are often represented in \emph{linear color spaces}, in which the pixel values are proportional to colorimetric, photometric or radiometric light quantities\footnote{The proportionality is approximate, as camera spectral sensitivity does not allow for direct measurement of colorimetric or photometric quantities.}. Notably, linear color spaces are very far from being perceptually uniform \cite{mantiuk2004perception}: the same change in physical units is much less visible at high brightness than low. Modern perceptually uniform transfer functions, such as the SMPTE PQ (perceptual quantizer) \cite{Miller2013} and PU21 (perceptually uniform transform) \cite{Mantiuk2021} have been developed to address this. By way of example, according to the PQ function, a change of 1\cdms{} starting from near darkness (0.005\cdms{}) is over 150 times more visible than the same change starting at 100\cdms{}. 

There are arguments both for and against training neural networks with images in linear spaces. If a plain L1 or L2 loss is used to train networks for linear data, we run the risk that the network will overfit for bright colors, and introduce unacceptable inaccuracies for darker colors. Additionally, if we wish to avoid visible quantization errors, pixel values in linear color spaces must be represented with more bits than display-encoded spaces. However, any physical phenomena involving light, such as lens blur, motion blur, or noise, can only be modeled in a physically plausible manner in linear color spaces. Therefore, physically plausible image formation models for blur, noise and sampling must be defined in linear color spaces, and by analogy, the inverse problems of deblurring, denoising and super-resolution should also be formulated in linear color spaces. It is unknown, however, whether neural networks can benefit from this connection to physical properties of linear color spaces.

In this work, we want to determine what training strategy leads to the best results. Should we train neural networks in display-encoded or linear color spaces (\secref{representations})? If we train networks in linear color spaces, should we use a loss function that compensates for perceptual non-uniformity (\secref{loss-functions})? 
There is no consensus in the literature on these questions. Many works train on images in linear color spaces and use an L1 loss \cite{Xu2020,Chen_Chen_Xu_Koltun_2018,conde_2024}. Some use a loss with perceptual encoding \cite{Eilertsen2017a,Kalantari2017,Mildenhall2022}. Our goal is to find empirical evidence to motivate a training strategy for image restoration techniques trained on HDR/RAW images. To that end, we collect datasets of HDR and RAW images and train six networks to perform common image restoration tasks: single-image super-resolution, denoising, and deblurring (\secref{testing}).

Rather than proposing a new method, this paper is a benchmark and reproduction study intended to settle these open questions. Our results serve as best practice guidelines for future ML methods operating on HDR and RAW content. Our results (\secref{results}) show a clear benefit in training networks in display-encoded color spaces, offering a gain in performance between 2 and 9\,dB for these applications.

\section{Related work}
\label{sec:rw}

% \subsection{Pixel Representations}
% \label{sec:rw_pixel_representations}

% \subsection{Loss Functions}
% \label{sec:rw_cost_functions}

% log - \cite{Eilertsen2017a}

% $\mu$-law \cite{Kalantari2017} - a range compressor used in audio processing

% PQ \cite{Miller2013}

% PU21 \cite{Mantiuk2021}

\subsection{Imaging RAW pixels}
% Done: Xiaoyu will add papers introducing it
% https://arxiv.org/pdf/2207.14671.pdf High Dynamic Range and Super-Resolution from Raw Image Bursts
% 
RAW image data, captured directly from image sensors, preserves unprocessed and uncompressed pixel data, thereby providing rich information for restoration techniques. RAW image formats also represent pixels in linear color spaces, which facilitate physically plausible modeling of camera noise \cite{Aguerrebere_2013}. Early works \cite{Foi2008,hwang2011difference} in this area mainly focused on noise reduction and artifact suppression based on statistical models of noise. With the advent of deep learning, restoration in the RAW domain has witnessed a paradigm shift. \cite{xie2012image, burger2012image, nam2016holistic} pioneered by the integration of deep neural networks for denoising. Those early works, however, targeted sRGB images, typically with synthetic additive Gaussian noise. Since denoising is likely to be integrated in the very early stages of the ISP pipeline, it should operate on linear RAW pixel values instead. 

Surprisingly, many deep-learning methods proposed for processing RAW images were never tested on RAW images \cite{Xing2021,Liu2020}. Instead, they opted to work with display-encoded images in the sRGB space, and assumed additive Gaussian noise. Both assumptions make input data very different from RAW sensor pixel values, raising the question on whether these methods would be equally effective for real camera data. Other methods that were trained and tested on RAW images \cite{Xu2020,Chen_Chen_Xu_Koltun_2018}, did so directly on linear color values and used an L1 loss. Our work demonstrates this strategy often fails to achieve good results. 
 %Instead, our experiments are performed on images represented in linear color space and using physically plausible camera noise model \cite{Foi2008}.

%These methods discussed so far, as well as most of the recent literature, use either L1 or L2 losses to train the networks. Xu et al. \cite{Xu2020} adopted an L1 loss and trained on linear color values for RAW image super-resolution. Xing and Egiazarian \cite{Xing2021} proposed to learn joint demosaicing, denoising, and super-resolution, and experimented on more loss functions including SSIM, MS-SSIM, and a mixture of those losses in the \RM{linear?} RGB color space.
%Liu et al. \cite{Liu2020} proposed an adaptive-threshold edge loss and edge-aware smoothness loss to recover the high-frequency details and smooth out the noise and unexpected artifacts of RGB images for the joint demosaicing and denoising task.

\subsection{High Dynamic Range (HDR) imaging}

The dynamic range is the contrast between the brightest and darkest parts of an image or video. Notably, the dynamic range the human eye can see is extremely high and much larger than what most standard dynamic range (SDR) capture and display pipelines are capable of. High Dynamic Range (HDR) is a term for a range of technologies that enhance the visual quality and realism of content by providing a wider than normal dynamic range \cite{reinhard2010high}. 

%High-dynamic-range (HDR) imaging endeavors to capture a wider range of luminance values than conventional imaging, thereby reproducing scenes of higher contrast. 

A key problem in HDR imaging involves merging multiple exposures into an HDR image while reducing ``ghosting'' artifacts due to motion \cite{Kalantari2017}. Another problem is tone-mapping of HDR scenes to the dynamic range that can be shown on a display \cite{reinhard2010high}. Yet another is the reconstruction of an HDR image from a single exposure (i.e. inverse-tone mapping) \cite{Eilertsen2017a,kim2019deep}. However, standard reconstruction problems like denoising, single-image super-resolution, or deconvolution/deblurring, are also relevant for HDR imaging. 

This work focuses on how HDR data in linear color spaces was handled in these previous works. Both Eilertsen et al. \cite{Eilertsen2017a} and Kalantari et al. \cite{Kalantari2017} trained the networks to directly predict HDR images in a linear color space. However, to compensate for the perceptual non-uniformity of that space,  Eilertsen et al. used the logarithmic function, and Kalantari et al. introduced the $\mu$-law (explained in detail in \secref{representations}). Mildenhall et al. \cite{Mildenhall2022} used a relative loss, similar to SMAPE (explained in \secref{loss-functions}), to train a NeRF on RAW images. The networks intended to reconstruct HDR UHD video content \cite{kim2019deep} were directly trained to predict display-encoded frames using the PQ OETF \cite{Miller2013}. 

%\RM{I checked \cite{chen2021hdrunet} --- using tanh function is a very bad idea and I would not like to advocate it by citing that paper}.

%Kim et al. \cite{kim2019deep} adopted an L2 loss in the YUV domain to conduct inverse tone-mapping and super-resolution for HDR applications. Going further, they employed the adversarial loss and a feature matching loss to the YUV frames \cite{kim2020jsi}.
%Several recent works aimed to obtain HDR images by combining multiple images with different exposures \cite{tan2021deep, chen2021hdrunet, chen2022attention, liu2023joint}. Notably, Chen and colleagues \cite{chen2021hdrunet} proposed the $Tanh$\_$L_1$ loss function to balance the impact of over-exposed values and well-exposed values. To avoid bright regions dominating an L1 loss for a linear HDR space, Liu et al. \cite{liu2023joint} applied the $\mu$-law tone-mapping function to the HDR image, and Mildenhall et al.~use a loss term weighted by the derivative of a hypothetical tone curve to reconstruct HDR views from a neural radiance field (NERF) \cite{Mildenhall2022}.

%\subsection{Perceptual Pixel Representations}
%\label{sec:perceptual_representations}

%introduced during bracketed-exposure merging processes \cite{}, enhancing the overall dynamic range and visual quality of SDR content \cite{}, or tone mapping \cite{}.

\section{Representations}
\label{sec:representations}

In this work, we explore popular perceptual pixel value representations from the literature for use in machine learning applications. The main question that arises is what representation --- linear or display encoded --- is most suitable for deep-learning networks. The linear representation should be better to model physical phenomena (assuming neural networks are able to take advantage of this benefit). However, the display-encoded representations are more data-efficient (requiring fewer bits) and have better perceptual uniformity. Moreover, most existing methods have been trained on display-encoded representations, as these are commonly used for SDR content widely available online. To answer the question above, we tested four representations: one linear and three display-encoded:
%\begin{itemize}
\paragraph{Linear:} linear RGB pixel values with BT.709 primaries, scaled between 0 and 1;

\paragraph{$\mu$-law:} a transfer function commonly used for perceptual audio encoding. Kalantari and Ramamoorthi \cite{Kalantari2017} proposed to encode HDR images using the $\mu$-law function: 
\begin{equation}
v_\mu(l)=\frac{\log(1+\mu\,l)}{\log(1+\mu)}
\label{eq:mu-law}
\end{equation}
with the constant $\mu=5000$. $l$ is a relative linear (RGB) color value scaled to the range 0--1. 

\paragraph{PQ (perceptual quantizer):}  display-encoding for HDR content \cite{Miller2013}, used in video coding and display standards (e.g., BT~2100). The PQ encoding was applied separately to each RGB channel in our experiments. Input linear values were given in absolute units (see \secref{training_cases}).

%The two display-encoded representations (PQ and PU21) require that the linear values that are encoded are scaled in absolute units (of light emitted from the display). We ensured that all input HDR content for these two representations is between 0.005 and 4000. This is the range of luminance that can be reproduced on high-quality HDR displays. The resulting display-encoded values were scaled between 0 and 1. 

\paragraph{PU21 (perceptually uniform transform):} a transform similar to PQ but based on more modern data. It was originally proposed as an encoding of HDR images to be used with existing standard dynamic range metrics \cite{Mantiuk2021,Aydn2008}.
For the PU21 encoding we use the quadratic function fitted to the original curve (which was derived numerically)
\begin{equation}
    v_\ind{PU21}(L) = a\,(\log_2(L)-L_\ind{min})^2 + b\,(\log_2(L)-L_\ind{min})\,,
    \label{eq:pu-enc}
\end{equation}    
where $L$ is an absolute linear (RGB) color value (in the range 0.005--10\,000) and the fitted parameters are $a = 0.001908$, $b = 0.0078$, and $L_\ind{min}=\log_2(0.005)$. The encoded values are between 0 and 1. The inverse is given by:
\begin{equation}
    v^{-1}_\ind{PU21}(V) = 2^{\frac{2\,a\,L_\ind{min} - b + \sqrt{b^2 + 4\,a\,V}}{2\,a}}\,.
\end{equation}
The functions above were fitted to the PU21 encoding for a banding model without the influence of glare (details in \cite{Mantiuk2021}) and are more computationally efficient than the PQ encoding or the original source code for PU21. This encoding was applied to each linear RGB channel in the same manner as PQ (see \secref{training_cases}). \\

%\end{itemize}

\begin{figure}
    \centering
    \includegraphics[width=\columnwidth]{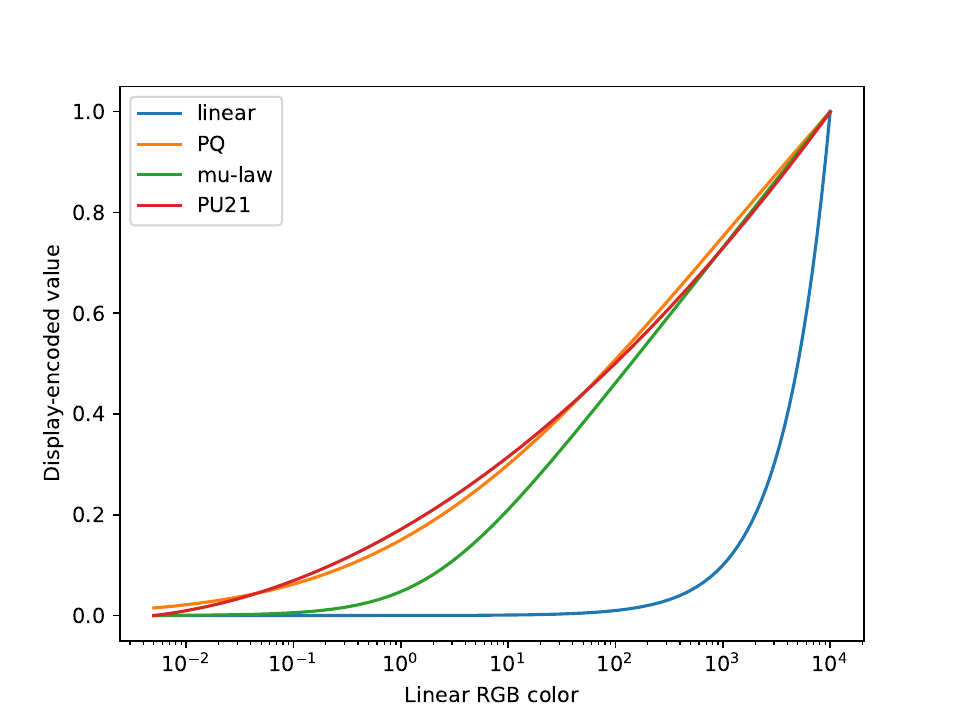}
    \caption{Perceptual encoding functions used to transform linear RGB color values to an approximately perceptually uniform space. A logarithmic scale is used for the x-axis to account for Weber's law. Following BT.2100, the range of interest for encoded values is between 0.005 and 10\,000. Note, however, that in our experiments, we used a smaller range between 0 and 4\,000. Since $\mu$-law encoding expects input values between 0 and 1, the values were divided by 10\,000 before passing to the function in \eqref{mu-law}. }
    \label{fig:transfer-functions}
\vspace{-3mm}
\end{figure}

The pixel encodings described above are plotted in \figref{transfer-functions}. It can be observed that both PQ and PU21 share very similar shapes, but PQ does not reach 0 for the minimum encoded value (0.005). The $mu$-law has a much shallower shape for values below 1, which means these values are encoded with less precision. Finally, linear encoding over-emphasizes large pixel values by allocating a very steep slope to those (note that the luminance is plotted on a logarithmic scale).

\section{Loss functions}
\label{sec:loss-functions}

In addition to the way pixels are encoded, the loss function is a key feature of training neural networks. Most image restoration problems can be effectively trained using L1 or L2 norms on pixel values as a loss function. However, this is not necessarily true when pixel values are represented in linear color spaces (e.g. RAW or HDR images), as they are perceptually non-uniform --- a small error at low brightness is more visible than the same error at large brightness levels. This problem can be avoided by first converting the compared pixel values to a color space that is more perceptually uniform so that the loss function becomes
\begin{equation}
    \ell=|| p(f_{\bm{\phi}}(\bm{x})) - p(\bm{y}) ||_1
\end{equation}
where $f_{\bm{\phi}}(\cdot)$ is a network with the weights $\bm{\phi}$, $\bm{x}$ is input, $\bm{y}$ is a reference (label) and the $p(\cdot)$ is a perceptual encoding function such as $\mu$-law, PQ or PU21 (all introduced in \secref{representations}). One important consideration is that the values passed to the function $p(\cdot)$ must be within the allowed range (e.g., greater than 0.005 for PQ and PU21). This means that out-of-range values must be clamped. Both PQ and PU21 require absolute photometric values, which correspond to light emitted from a display (e.g., between 0.005 and 10\,000). 

The final loss we investigate is the symmetric mean absolute percentage error (SMAPE):
\begin{equation}
    \ell = \norm{\frac{|f_{\bm{\phi}}(\bm{x})-\bm{y}|}{|f_{\bm{\phi}}(\bm{x})| + |\bm{y}| + \epsilon} }_1\,,
\end{equation}
where $\epsilon$ is a small constant avoiding division by 0. Dividing by the sum of absolute values makes the error relative and more perceptually uniform, following Weber's law. 
\section{Experiments}
\label{sec:testing}

In this work, we focus on image restoration problems (denoising, deblurring, and single-image super-resolution), which aim to generate a higher-quality version of the input image. We exclude higher-level tasks (e.g., recognition) as they involve very different losses to those used for restoration. We also do not consider the tasks in which input and output images differ in dynamic range --- tone mapping and single-image HDR reconstruction --- as these tasks would only allow us to test either the representations (for tone mapping) or the loss functions (single-image HDR reconstruction) but not both at once. 

To compare the effectiveness of different loss functions and pixel encodings, we train networks with HDR images on three tasks: image denoising, deblurring, and super-resolution. For each task, we choose two different neural networks: one a well-established technique and the second a more recent method. We follow the training settings proposed in the original works. EDSR (2017) \cite{lim2017enhanced} and Real-ESRGAN (2021) \cite{wang2021real} are used for the task of single-image super-resolution (4x). The downsampled images used for training were created using a bilinear filter.  For image deblurring, we train GFNNet (2018) \cite{zhang2018gated} and MirNet-v2 (2022) \cite{zamir2022learning}. The networks are trained on images affected by Gaussian blur with sigma values of 8\,px in both directions. DnCNN (2017) \cite{zhang2017beyond} and SADNet (2020) \cite{chang2020spatial} are used for denoising of HDR images. All the tested methods were originally designed for display-encoded (SDR) images. Therefore, our tests demonstrate how well they generalize to HDR data and what is the best training strategy. 

%% OBS.: (ALEX) I took a quick look at Lei's code for the noise model, and it was not immediately apparent how this could be written down here. I am skipping this task and, for now assume that the provided reference covers the model well enough for the purposes of this text.

%The formulation of the noise model is provided in the \RM{ supplementary} \alex{Should we just add it here? The paper is relatively short, so we can just add a subsection/paragraph for it}.  \lei{formula is somewhat long} \RM{I am sure we can fit it in the main paper.}

All networks are trained using the code provided by the authors on their GitHub repositories. We adopted the same training schemes as the original papers to train each network, including the optimizer and learning rate. The exception was Real-ESRGAN for which we used only L1 loss (disabled the perceptual loss) and adjusted the learning rates as listed in \tableref{real-ESRGAN-lr} (in supplementary). We trained each network for a different number of epochs to ensure that convergence was reached. EDSR was trained for 3k epochs, 150 epochs for Real-ESRGAN, 1k epochs for MirNet-v2, GFNNet, DnCNN, and SADNet.

%The training for super-resolution methods was completed on NVidia A100 GPUs and for on \RM{To check} for other methods. 

\subsection{Datasets}
\label{sec:dataset}

We used two datasets: one for testing reconstruction on HDR images and another for testing on camera RAW images. The first dataset consisted of 106 HDR images from the HDR Photographic Survey \cite{fairchild_hdr_2007} and HDR video frames from the SJTU HDR Video Sequences \cite{Song2016TheSH}. Since frames originating from the same video share similarities, we used one representative HDR frame provided for each video\footnote{The frames can be downloaded from the  \href{https://medialab.sjtu.edu.cn/files/SJTU\%20HDR\%20Video\%20Sequences/demo_images/}{website}.}, adding an additional 16 reference images to the dataset. 60\% of images were selected for training, 20\% for validation, and the remaining 20\% were used for testing. To verify the networks' generalization ability to exposure variations, the testing data set was augmented five-fold by multiplying it with a random exposure coefficient drawn from a uniform distribution between 0.1 and 0.9. Physically plausible simulated camera noise (photon and readout noise) was added to training images \cite{Foi2008}. 

Additionally, to show that our observations generalize to RAW images, we test super-resolution on the \emph{Learning to See in the Dark} dataset \cite{Chen_Chen_Xu_Koltun_2018}. We utilized 231 long-exposure RAW images from this dataset \cite{Chen_Chen_Xu_Koltun_2018}, with the same 60/20/20 split for training, validation, and testing. The color values were reconstructed from RAW values using Adaptive Homogeneity-Directed (AHD) demosaicing \cite{Hirakawa_Parks_2005} and converted to the BT.709 color space (linear). The demosaicing step was necessary as the tested architectures expected RGB images as input. Because RAW pixel values are relative, we had to rescale RGB values to the absolute range suitable for the PU21 and PQ encoding. We adjusted exposure of each image so that the average luminance was 20 nits. Such exposure adjustment matched the average luminance of SDR content and made images sufficiently bright without saturating the brightest pixels. The histograms of pixel values across both datasets can be found \figref{dataset-histograms} of the supplementary document. Note that due to the computational constraints, we could test only super-resolution methods on the RAW dataset. 

%To make the network robust to exposure variations, \lei{
%We did not directly use RAW images in our experiments because they tend to contain more noise than their HDR counterparts (merging multiple exposures can radically reduce noise). However, the simulated noise used for denoising and the exposure augmentation technique made the prepared HDR images very close to demosaiced RAW images. 

\subsection{HDR/RAW training cases}
\label{sec:training_cases}

\begin{table}
    \caption{The eight combinations of pixel encoding and loss functions tested in our experiments.}
    \centering
    {\small
    \begin{tabular}{lcc}
    \toprule
         Label & Pixel encoding & Loss function \\
         \midrule
         \condition{Linear-L1} & Linear & L1 \\
         \condition{PQ-L1} & PQ & L1 \\
         \condition{PU21-L1} & PU21 & L1 \\
         \condition{$\mu$-L1} & $\mu$-law (mu-law) & L1 \\
         \condition{Linear-PQ} & Linear & PQ \\
         \condition{Linear-PU21} & Linear & PU21 \\
         \condition{Linear-$\mu$} & Linear & $\mu$-law (mu-law) \\
         \condition{Linear-SMAPE} & Linear & SMAPE \\
         \bottomrule
    \end{tabular}
    }
    \label{tab:tested-conditions}
\end{table}

\begin{figure}
   \centering
   \includegraphics[width=\columnwidth]{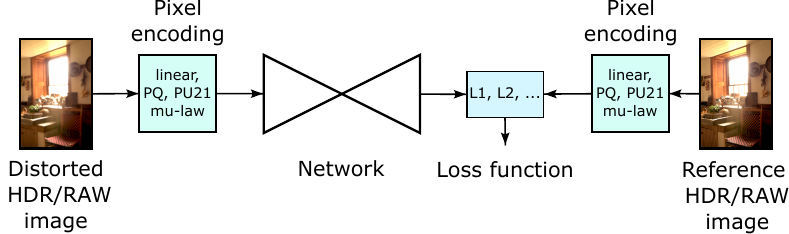}
   \caption{Training on HDR/RAW images. Training pairs were first (optionally) encoded from a linear to a perceptually uniform color space using PQ, PU21 or a $\mu$-law transfer function (explained in \secref{representations}). If no pixel encoding was applied (linear), one of the loss functions from \secref{loss-functions} was used to account for the perceptual non-uniformity. }
    \vspace{-4mm}
   \label{fig:training-diagram}
\end{figure}

We tested 8 combinations of pixel encodings and loss functions, listed in \tableref{tested-conditions}. All perceptual pixel encodings (PQ, PU21, and $\mu$-law, see \secref{representations}) were used with a regular L1 loss because the encoded values are expected to be approximately perceptually uniform and, therefore, should not require a custom loss function. However, when no encoding was used (labeled \emph{linear}), we used either an L1 loss or one of the losses described in \secref{loss-functions}. We also experimented with losses that were the sum of two terms, such as L1 and PQ-encoded L1, but they performed no better than individual loss terms.  \figref{training-diagram} shows a diagram of our training setup. 

Since PQ and PU21 both require pixel values in absolute units (of light, as emitted by a display), the linear pixel values were scaled to the range 0.005--4\,000. Although both encodings allow values up to 10\,000 (nit), 4\,000 was selected as being more representative of realistic peak luminance values for a high-quality HDR display. For other pixel encodings, the linear color values were scaled in the 0--1 range. By following these steps, we ensure all values passed to the networks (for every encoding) were between 0 and 1, as is typical for training.  

% \begin{figure*}
%     \centering
%     \includegraphics[width=0.49\linewidth]{results/real-psnr.pdf}    
%     \includegraphics[width=0.49\linewidth]{results/edsr-psnr.pdf} \\    
%     \includegraphics[width=0.49\linewidth]{results/real-cvvdp.pdf}    
%     \includegraphics[width=0.49\linewidth]{results/edsr-cvvdp.pdf} \\
%     \includegraphics[width=0.49\linewidth]{results/real-ssim.pdf}    
%     \includegraphics[width=0.49\linewidth]{results/edsr-ssim.pdf} 
%     \caption{Single-image super-resolution results for the two networks (columns) and three metrics (rows). The violin shape visualizes the distribution across the testing set. The thin black line shows the range of quality scores (excluding outliers), and the thicker line shows the region between the 25th and 75th percentiles. The dot and the numerical value represent the median. The red horizontal lines above the violin plots denote the groups of conditions for which we have no evidence of statistically significant differences. The results are sorted from the best on the left to the worst on the right. The colors are different for each combination of the representation and a loss and are consistent across the plots.}
%     \label{fig:sisr-results}
% \end{figure*}

\begin{figure*}
    \centering
    \includegraphics[width=0.95\linewidth]{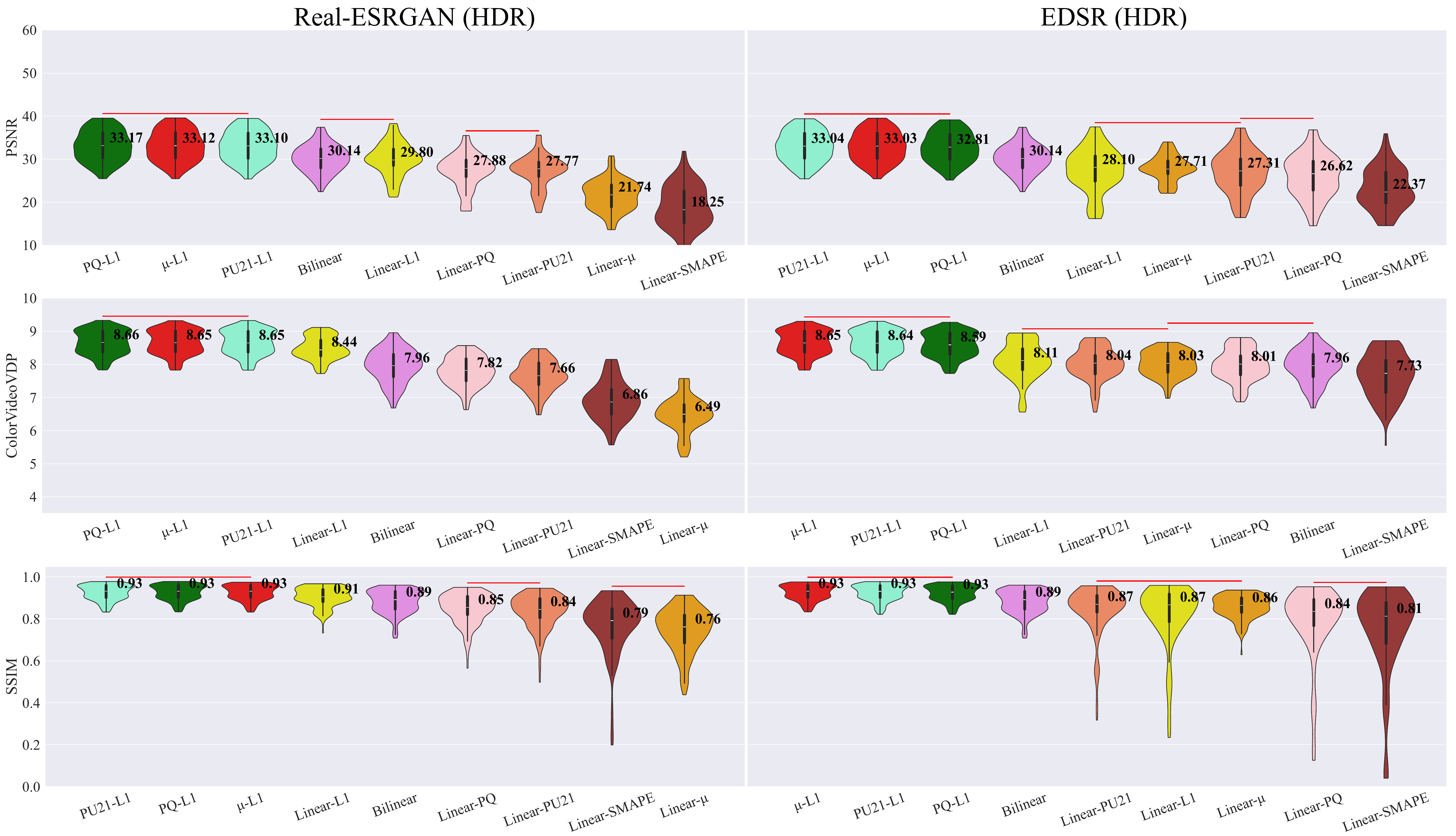} \\
    \includegraphics[width=0.95\linewidth]{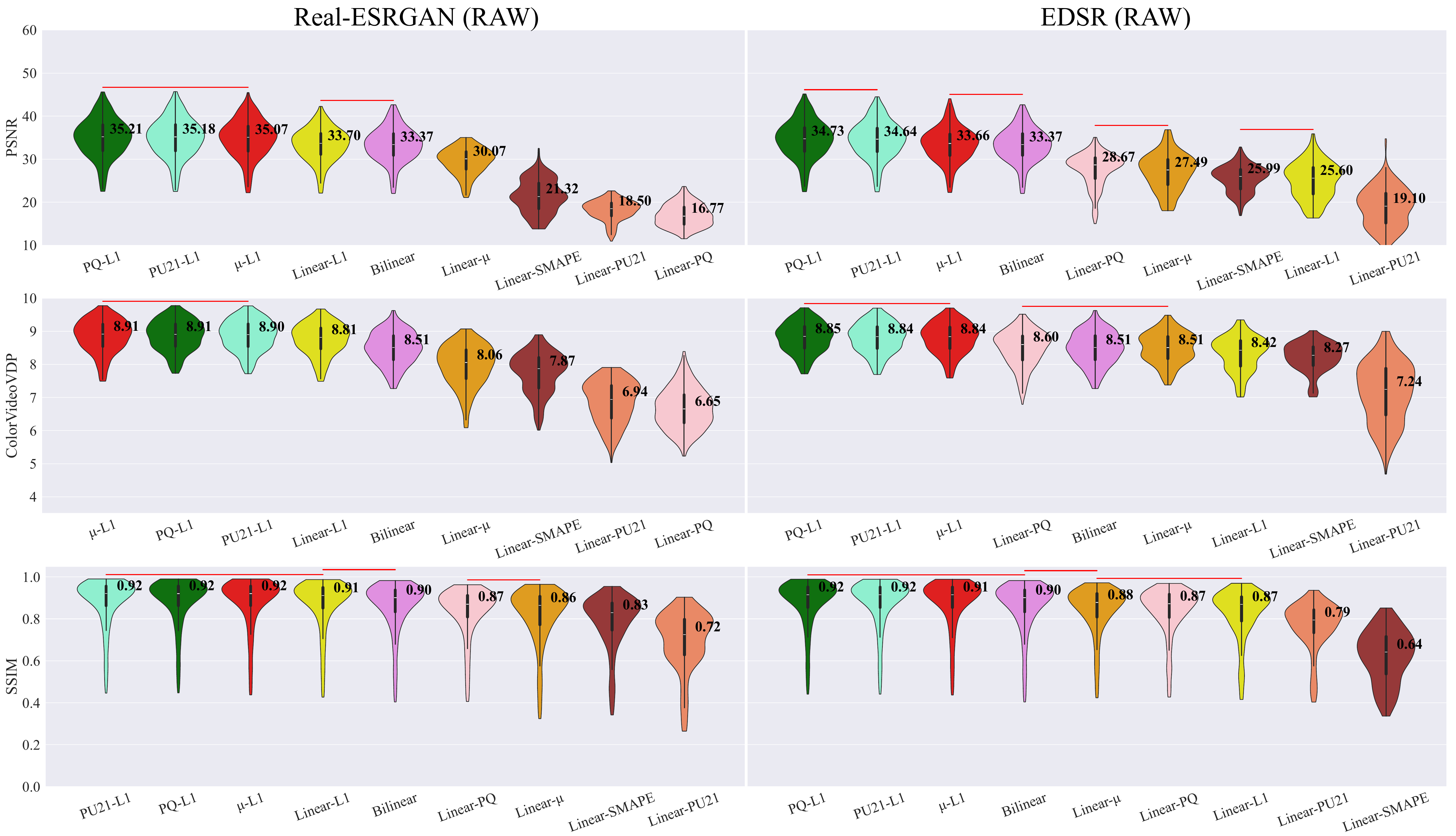}    
    \caption{Single-image super-resolution results for the HDR (top) and RAW (bottom) datasets, the two networks (Real-ESRGAN and EDSR), and three metrics. The violin shape represents the distribution across the testing set. A thin black line shows the range of quality scores (excluding outliers), and the thicker line shows the region between the 25th and 75th percentiles. The dot and numerical value represent the median. Red horizontal lines above the violin plots denote the groups of conditions for which we have no evidence of statistically significant differences. Results are sorted from the best on the left to the worst on the right. The colors are different for each combination of representation and loss and are consistent across plots.}
    \label{fig:sisr-results}
\end{figure*}

\section{Results}
\label{sec:results}

To evaluate the results, we used three image quality metrics: PSNR, SSIM and ColorVideoVDP \cite{Mantiuk_2024_cvvdp}. PSNR and SSIM were selected as the two most widely used metrics. ColorVideoVDP was selected because it natively supports HDR images and was designed to compare color images and videos. To adapt PSNR and SSIM to HDR images, we used the PU21 encoding \cite{Mantiuk2021}, i.e. PSNR was computed on the PU21-encoded RGB pixels. SSIM was computed on the luma channel of the PU21-encoded pixels.

\subsection{Single-image super-resolution}

The numerical results for the two single-image super-resolution methods can be found in \figref{sisr-results} and in the top row of \tableref{results}, and examples of visual results in \figref{sisr-vis-res-hdr} (in supplementary) for HDR and in \figref{sisr-vis-res-raw} (in supplementary) for the RAW dataset. In \figref{sisr-results}, we visualize the distribution of the quality scores across the entire test set and also report the results of pair-wise t-tests (two-tailed, $\alpha=0.05$, $N=144$ for HDR (24 test images $\times$ 6 exposures) and $N=276$ for RAW (46 test images $\times$ 6 exposures)). The violin plots are sorted left-to-right from the best-performing condition to the worst. Red horizontal lines connecting two or more violin plots signify no evidence of statistically significant differences within that group. Conversely, conditions that are not joined by a common line were found to have performances that differ at a statistically significant level.

% Instead of providing a table showing only mean results, 

The results in \figref{sisr-results} indicate a substantial gain in performance of both super-resolution methods (1.5--9\,dB, as compared to \condition{Linear-L1}) when one of the perceptual pixel representations is used. However, we have no statistical evidence showing that one perceptual representation is better than the others. Interestingly, except for the ESDR-RAW experiment, perceptual loss functions did not improve the results of either super-resolution methods with respect to the ``plain'' \condition{Linear-L1}. The visual results in \figref{sisr-vis-res-raw} and \figref{sisr-vis-res-hdr} show that networks trained on \condition{Linear} representations often result in color artifacts. The networks trained on perceptual representations (\condition{PU-L1}, \condition{PQ-L1} and \condition{$\mu$-L1}) do not suffer from those artifacts and produce sharper images. 

% To further verify the performance of networks trained with different pixel encodings and loss functions, we conducted another evaluation on a blind dataset (MIT-Adobe 5K Dataset), on which the networks were not trained. Results in \tableref{results-blind} show that networks trained with  $\mu$-law or PU21 pixel encodings still perform better than other configurations.

% \begin{figure*}
%     \centering
%     \includegraphics[width=0.37\linewidth]{results/gfnnet-psnr.pdf}
%     \includegraphics[width=0.61\linewidth]{results/mir-psnr.pdf} \\
%     \includegraphics[width=0.37\linewidth]{results/gfnnet-cvvdp.pdf}
%     \includegraphics[width=0.61\linewidth]{results/mir-cvvdp.pdf} \\
%     \includegraphics[width=0.37\linewidth]{results/gfnnet-ssim.pdf}
%     \includegraphics[width=0.61\linewidth]{results/mir-ssim.pdf} 
%     \caption{Debluring results for the two networks (columns) and three metrics (rows). A few configurations are missing for GFNet as network training failed to converge for those. The notation is the same as in \figref{sisr-results}.}
%     \label{fig:deblur-results}
% \end{figure*}

\begin{table*}[ht]
\caption{The median values for evaluation results. The last column ``Naive'' is bilinear-upsampling results for super-resolution experiments, original blurry images for deblurring experiments, and the original noisy images for denoising experiments. The best-performing scores are annotated in red, while the second-best is in blue. The "N/A" entries correspond to the cases for which the training failed.}
\centering
\resizebox{\textwidth}{!}{
\begin{tabular}{lcllllllllll}
\hline
\multicolumn{3}{l}{} & PU21-L1 & PQ-L1 & $\mu$-L1 & Linear-L1 & Linear-PU21 & Linear-PQ & Linear-$\mu$ & Linear-SMAPE & Naive \\
\cline{2-12}
\multirow{6}{*}{\begin{tabular}[c]{@{}l@{}}{\rotatebox[origin=c]{90}{SI-SR --- HDR}}\end{tabular}} & \multirow{3}{*}{EDSR} & SSIM & {\color[HTML]{FE0000} 0.93}& {\color[HTML]{FE0000} 0.93}& {\color[HTML]{FE0000} 0.93}& 0.87& 0.87& 0.84& 0.86& 0.81& {\color[HTML]{3166FF} 0.89}\\
 &  & PSNR & {\color[HTML]{FE0000} 33.04}& 32.81& {\color[HTML]{3166FF}33.03 }& 28.10& 27.31& 26.62& 27.71& 22.37& 30.14\\
 &  & CVVDP & {\color[HTML]{3166FF} 8.64}& 8.59& {\color[HTML]{FE0000} 8.65}& 8.11& 8.04& 8.01& 8.03& 7.73& 7.96\\
 \cline{2-12}
 & \multicolumn{1}{l}{\multirow{3}{*}{\begin{tabular}[c]{@{}l@{}}{Real-ESRGAN}\end{tabular}}} & SSIM & {\color[HTML]{FE0000} 0.93}& {\color[HTML]{FE0000} 0.93}& {\color[HTML]{FE0000} 0.93}& {\color[HTML]{3166FF} 0.91}& 0.84& 0.85& 0.76& 0.79& 0.89\\
 & \multicolumn{1}{l}{} & PSNR & 33.10& {\color[HTML]{FE0000} 33.17}& {\color[HTML]{3166FF} 33.12}& 29.80& 27.77& 27.88& 21.74& 18.25& 30.14\\
 & \multicolumn{1}{l}{} & CVVDP & {\color[HTML]{3166FF} 8.65}& {\color[HTML]{FE0000} 8.66}& {\color[HTML]{3166FF} 8.65}& 8.44& 7.66& 7.82& 6.49& 6.86& 7.96\\
 \cline{2-12}
\multirow{6}{*}{\begin{tabular}[c]{@{}l@{}}{\rotatebox[origin=c]{90}{SI-SR --- RAW}}\end{tabular}}& \multirow{3}{*}{EDSR} & SSIM & {\color[HTML]{FE0000} 0.92}& {\color[HTML]{FE0000} 0.92}& {\color[HTML]{3166FF} 0.91}& 0.87& 0.79& 0.87& 0.88& 0.64& 0.90\\
 &  & PSNR & {\color[HTML]{3166FF} 34.64}& {\color[HTML]{FE0000} 34.73}& 33.66& 25.60& 19.10& 28.67& 27.49& 25.99& 33.37\\
 &  & CVVDP & {\color[HTML]{3166FF} 8.84}& {\color[HTML]{FE0000} 8.85}& {\color[HTML]{3166FF} 8.84}& 8.42& 7.24& 8.60& 8.51& 8.27& 8.51\\
 \cline{2-12}
 & \multicolumn{1}{l}{\multirow{3}{*}{\begin{tabular}[c]{@{}l@{}}{Real-ESRGAN}\end{tabular}}} & SSIM & {\color[HTML]{FE0000} 0.92}& {\color[HTML]{FE0000} 0.92}& {\color[HTML]{FE0000} 0.92}& {\color[HTML]{3166FF} 0.91}& 0.72& 0.87& 0.86& 0.83& 0.90\\
 & \multicolumn{1}{l}{} & PSNR & {\color[HTML]{3166FF} 35.18}& {\color[HTML]{FE0000} 35.21}& 35.07& 33.70& 18.50& 16.77& 30.07& 21.32& 33.37\\
 & \multicolumn{1}{l}{} & CVVDP & {\color[HTML]{3166FF} 8.90}& {\color[HTML]{FE0000} 8.91}& {\color[HTML]{FE0000} 8.91}& 8.81& 6.94& 6.65& 8.06& 7.87& 8.51\\
\hline
\multirow{6}{*}{\rotatebox[origin=c]{90}{Deblurring}} & \multirow{3}{*}{GFNNet} & SSIM & {\color[HTML]{FE0000} 0.94} & {\color[HTML]{3166FF} 0.93} & 0.92 & 0.83 & N/A & N/A & 0.88 & N/A & 0.87  \\
 &  & PSNR & {\color[HTML]{3166FF} 33.52} & 32.57 & {\color[HTML]{FE0000} 33.65} & 29.98 & N/A & N/A & 32.80 & N/A & 27.30 \\
 &  & CVVDP & {\color[HTML]{3166FF} 7.59} & 7.27 & {\color[HTML]{FE0000} 7.66} & 7.09 & N/A & N/A & 7.41 & N/A & 5.51 \\
 \cline{2-12}
 & \multirow{3}{*}{MirNet-v2} & SSIM & 0.93 & {\color[HTML]{FE0000} 0.95} & {\color[HTML]{3166FF} 0.94} & 0.93 & 0.93 & {\color[HTML]{FE0000} 0.95} & {\color[HTML]{FE0000} 0.95} & {\color[HTML]{FE0000} 0.95} & 0.87  \\
 &  & PSNR & 32.99 & {\color[HTML]{3166FF} 34.65} & 34.10 & 31.98 & 32.06 & 33.57 & {\color[HTML]{FE0000} 34.80} & 34.10 & 27.30 \\
 &  & CVVDP & 7.40 & {\color[HTML]{3166FF} 7.89} & 7.77 & 7.31 & 7.35 & 7.73 & {\color[HTML]{FE0000} 7.93} & 7.72 & 5.51 \\
\hline
\multirow{6}{*}{\rotatebox[origin=c]{90}{Denoising}} & \multirow{3}{*}{DnCNN} & SSIM & {\color[HTML]{3166FF} 0.85} & 0.84 & {\color[HTML]{FE0000} 0.86} & 0.80 & 0.83 & 0.83 & 0.82 & 0.80 & 0.24  \\
 &  & PSNR & {\color[HTML]{3166FF} 23.10} & 22.44 & {\color[HTML]{FE0000} 23.65} & 17.93 & 20.56 & 21.03 & 20.56 & 18.57 & 10.60 \\
 &  & CVVDP & {\color[HTML]{3166FF} 6.07} & 5.96 & {\color[HTML]{FE0000} 6.15} & 5.68 & 5.87 & 5.74 & 5.76 & 5.66 & 4.37  \\
 \cline{2-12}
 & \multirow{3}{*}{SADNet} & SSIM & {\color[HTML]{3166FF} 0.90} & {\color[HTML]{FE0000} 0.91} & {\color[HTML]{3166FF} 0.90} & 0.88 & 0.89 & 0.89 & 0.88 & 0.89 & 0.24  \\
 &  & PSNR & 23.94 & {\color[HTML]{FE0000} 27.76} & {\color[HTML]{3166FF} 26.47} & 24.62 & 26.03 & 25.32 & 26.32 & 26.08 & 10.60 \\
 &  & CVVDP & 6.52 & {\color[HTML]{FE0000} 6.86} & 6.61 & 6.71 & {\color[HTML]{3166FF} 6.73} & 6.68 & 6.67 & 6.71 & 4.37 \\
 \hline
\end{tabular}
}
\label{tab:results}
\end{table*}

\subsection{Deblurring}

The deblurring results are different for the two tested methods and will be discussed separately. GFNNet failed to train and converge for several of the tested loss functions; it did, however, converge for all perceptual pixel encodings. As shown in \figref{deblur-results} and \tableref{results}, the best performance was obtained using either the $\mu$-law or PU21 pixel encodings, resulting in an improvement of about 3.8\,dB compared to the plain \condition{Linear-L1} configuration. The improvement was smaller (but still significant) for the PQ pixel encoding and linear encoding using the $\mu$-law loss function. Visually (see \figref{gfnet-vis-res} in supplementary) the \condition{Linear-L1} configuration resulted in artifacts in the form of vertical stripes. These artifacts were also present but less noticeable for other configurations. The \condition{PQ-L1} configuration resulted in a slightly softer image. 

Deblurring with MirNet-v2 produced better results than for GFNNet. Similarly as for GFNNet, the worst results were achieved by the plain \condition{Linear-L1} configuration. The best results were obtained by a linear encoding with the $\mu$-law loss, PQ and $\mu$-law pixel encodings (gain of 2--3\,dB). Interestingly, the PU21 pixel encoding resulted in worse results (see \figref{mirnet-vis-res} in supplementary, note the visual results' softer appearance), though the reason is unclear. %The visual results in \figref{mirnet-vis-res} show that the network trained with the PU21 pixel encoding produced a softer result. 

\begin{figure*}[t]
    \centering
    \includegraphics[width=0.90\textwidth]{results/deblur_noTransparency.pdf}
    \caption{Deblurring results for the two networks (columns) and three metrics (rows). A few configurations are missing for GFNet as network training failed to converge for those. The notation is the same as in \figref{sisr-results}.}
    \label{fig:deblur-results}
\end{figure*}

% \begin{figure*}
%     \centering
%     \includegraphics[width=0.49\linewidth]{results/dncnn-psnr.pdf}
%     \includegraphics[width=0.49\linewidth]{results/sad-psnr.pdf} \\
%     \includegraphics[width=0.49\linewidth]{results/dncnn-cvvdp.pdf}
%     \includegraphics[width=0.49\linewidth]{results/sad-cvvdp.pdf}  \\
%     \includegraphics[width=0.49\linewidth]{results/dncnn-ssim.pdf}
%     \includegraphics[width=0.49\linewidth]{results/sad-ssim.pdf} 
%     \caption{Denoising results for the two networks (columns) and three metrics (rows) The notation is the same as in \figref{sisr-results}.}
%     \label{fig:denoise-results}
% \end{figure*}

\begin{figure*}[t]
    \centering
    \includegraphics[width=0.90\textwidth]{results/denoise_noTransparency.pdf}
    \caption{Denoising results for the two networks (columns) and three metrics (rows) The notation is the same as in \figref{sisr-results}.}
    \label{fig:denoise-results}
\end{figure*}

\subsection{Denoising}

For denoising networks, PSNR had more distinctive and slightly different results than ColorVideoVDP and SSIM, as seen in \figref{denoise-results} and \tableref{results}. We investigated this discrepancy further by comparing pairs of conditions for which the metric predictions differed the most. We observed that both denoising methods, in particular DnCNN, resulted in images that differed in brightness, tone, color, and contrast from the original. Even though such changes may be less perceptually objectionable (and difficult to notice without a reference), PSNR was very sensitive to them. Both ColorVideoVDP and SSIM put more emphasis on structural distortions and seemed to correlate better with the perceived quality. 

%However, SSIM was very sensitive to noise, even when it was barely noticeable. Therefore, our analysis is focused on the ColorVideoVDP predictions, which seem to be best aligned with the perceived image quality. 

For both denoising methods, perceptual pixel encodings produced the best results in most cases. $\mu$-law and PU21 performed the best for DnCNN, followed by PQ, which performed slightly worse according to PSNR and ColorVideoVDP. In the case of SADNet, ColorVideoVDP could not distinguish between any of the configurations (no evidence of statistical differences). While $\mu$-law consistently produced the highest metric scores for SADNet, PU21 was the second best in terms of SSIM, but the last in terms of PSNR. We attribute this discrepancy to color shifts (SSIM is insensitive to color changes). The visual results (see \figref{dncnn-vis-res} in supplementary) show changes in color for Linear-PQ and Linear-PU. As a result, these two configurations have lower PSNR than the input noisy image (but higher ColourVideoVDP index because of the denoising).

\section{Discussion and conclusions}

The results across six tested methods clearly indicate that networks for image restoration should be trained on HDR/RAW images that are encoded using one of the perceptual transforms: PU21, PQ, or $\mu$-law. We did not find sufficient evidence to recommend one transform over another across all applications. However, we observed slightly worse performance of PQ for GFNNet, and PU21 for MirNet-v2 and SADNet. $\mu$-law encoding was robust across different methods but did not always produce the best results. Both $\mu$-law and PU21 use simpler and more computationally efficient formulas. PQ fails to scale the resulting values to a 0--1 range (see \figref{transfer-functions}), which may cause practical issues. Unlike the $\mu$-law, PQ and PU21 were derived from psychophysical data and have stronger perceptual bases. As the differences in performance between the three transforms are mostly small, we recommend using any of them, as this offers a substantial gain in performance over not using them at all. The actual gain in performance is application-specific and can be as large as 2--9\,dB as compared to the ``default'' approach of training linear color values with an L1 loss. Using linear color values in tandem with a loss function that employs perceptual encoding seems to be less effective across the tested applications. 

Although we cannot provide a conclusive answer of why perceptual transforms are beneficial for training neural networks, we suspect that this is because they offer better data efficiency. Pixels encoded with PU21, PQ, or $\mu$-law require fewer bits to represent the same amount of visually relevant information. Although networks should be able to learn such a transform during training, this removes some capacity from the network, which could otherwise be devoted to their main task. Some works avoid the conundrum of what representation to use by providing both linear and gamma-encoded images to the network \cite{Kalantari2017,Yan_2019}. This, however, leads to larger networks, reduced capacity or efficiency of the trained models.

%\alex{The arguments in this paragraph are solid, but if I was a practitioner, I would still want to see a clear conclusion of what I should use, even if it doesn't have statistical significance. Can we do an average dB gain over all examined applications and pick 1 winner to recommend? I did a simple test and found the mean of the PSNRs across all applications for the 3 encoding methods, and got PU=30.895, PQ=31.5533, Mu=31.5283. Would it be OK to recommend PQ, but note that the performance, while generally good, may not be optimal for every application?} \RM{Well, if we have no strong evidence, we should not promote one method over another. The recommendation would be to use one of them, or try to use all of them.}

The empirical findings in this work are relevant as they show that very small changes in training strategy (``one-liners'') can substantially improve the results of neural networks trained for popular tasks on HDR or RAW images. Moreover, the winning strategy of using perceptual pixel coding is rarely used in practice. More often, networks are trained in linear color spaces using an L1 cost function \cite{Xu2020,Chen_Chen_Xu_Koltun_2018,conde_2024}, or at most, use a perceptual encoding loss \cite{Eilertsen2017a,Kalantari2017,Mildenhall2022} --- both strategies shown in this work to under-perform. We hope our findings help practitioners in designing better training strategies for HDR/RAW images.

{
    \small
    \bibliographystyle{ieeenat_fullname}
    \bibliography{bibliography}
}

\renewcommand{\thesection}{S\arabic{section}} 
\renewcommand{\thefigure}{S\arabic{figure}}
\renewcommand{\theequation}{S\arabic{equation}}
\renewcommand{\thetable}{S\arabic{table}}

% \begin{abstract}
%   This is a supplementary material for \textquotesingle \textquotesingle. 
% \end{abstract}


\clearpage
\setcounter{page}{1}
\maketitlesupplementary

\setcounter{figure}{0}
\setcounter{section}{0}
\setcounter{equation}{0}
\setcounter{table}{0}

% <concept_id>10010147.10010257.10010258.10010259.10010266</concept_id>
% <concept_desc>Computing methodologies~Cost-sensitive learning</concept_desc>
% <concept_id>10010147.10010371.10010382.10010236</concept_id>
% <concept_desc>Computing methodologies~Computational photography</concept_desc>
% <concept_significance>300</concept_significance>

% \ccsdesc[500]{Computing methodologies~Cost-sensitive learning}
% \ccsdesc[300]{Computing methodologies~Computational photography}

% \keywords{HDR, loss function, image reconstruction}

% \citestyle{acmauthoryear}

%% Use this when submitting an article to a sponsored event. You'll
%% receive a unique submission ID from the organizers
%% of the event, and this ID should be used as the parameter to this command.

% \acmSubmissionID{papers\_375}


% \begin{document}

% \maketitle

\section{Content}

\begin{figure}[t]
    \centering
    \includegraphics[width=0.4\textwidth]{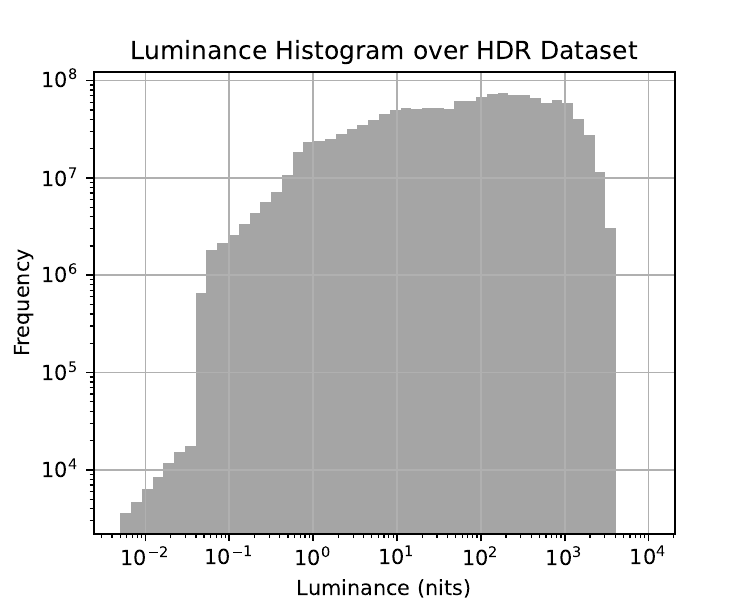}
    \includegraphics[width=0.4\textwidth]{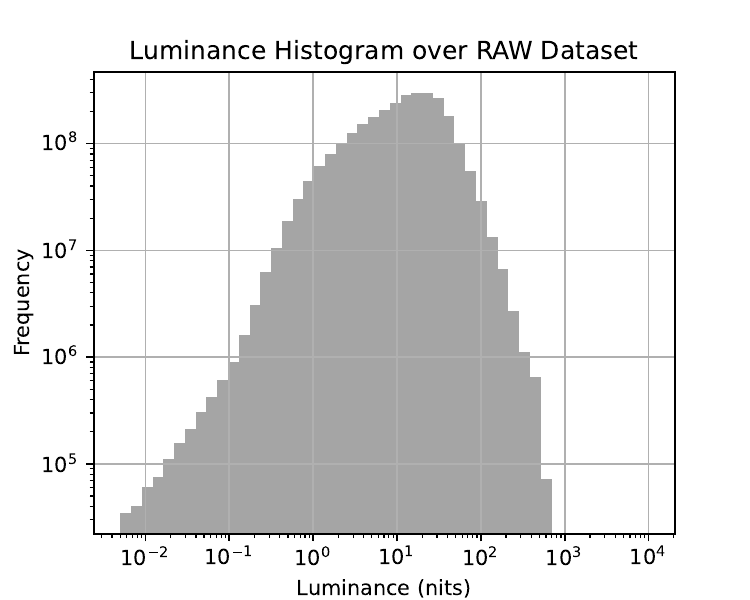}
    \caption{Histogram over all images in the HDR (top) and RAW (bottom) datasets.}
    \label{fig:dataset-histograms}
\end{figure}

\begin{figure*}
    \centering
    \includegraphics[width=1\linewidth]{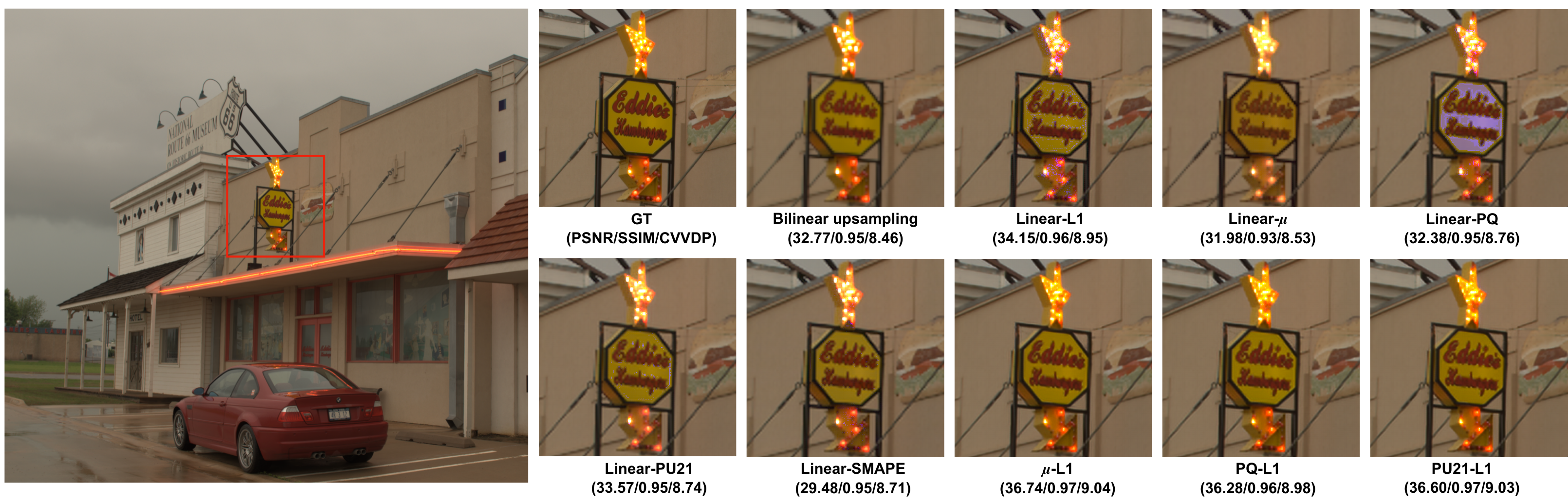} \\
    \includegraphics[width=1\textwidth]{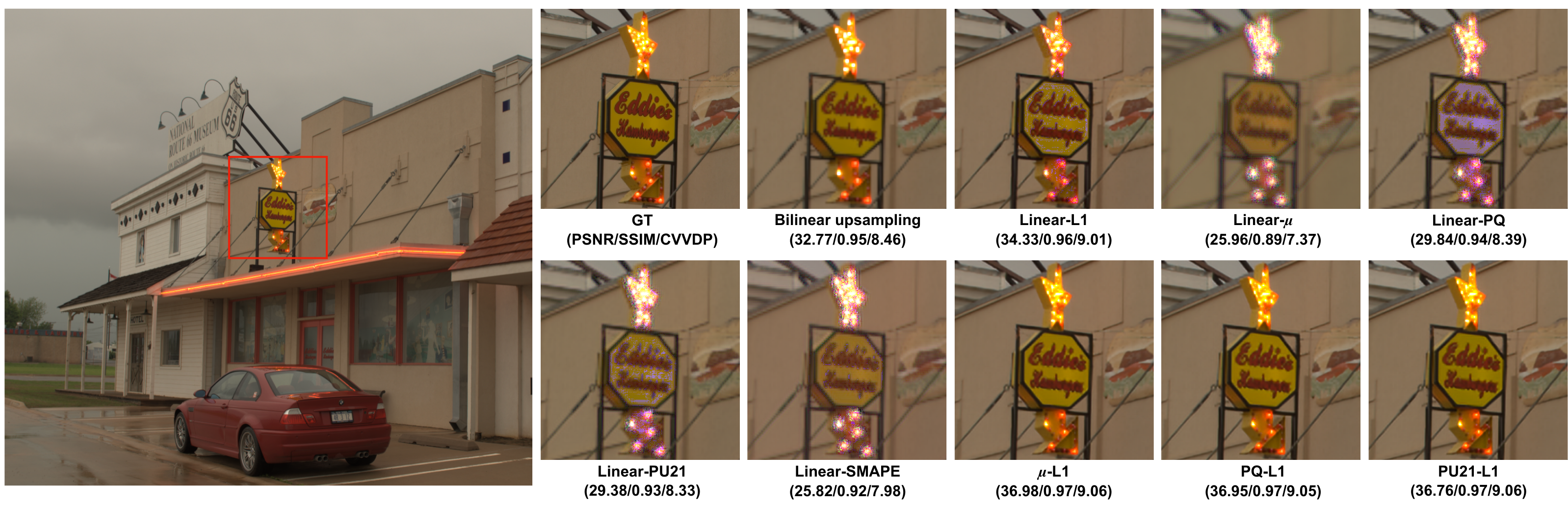}
    \caption{Example results for single-image super-resolution on HDR images with EDSR  \cite{lim2017enhanced} (top) and Real-ESRGAN \cite{wang2021real} (bottom). The numbers in parentheses show PSNR, SSIM, and ColorVideoVDP quality values (the higher, the better) for the reconstructed image.}
    \label{fig:sisr-vis-res-hdr}
\end{figure*}

\begin{figure*}
    \centering
    \includegraphics[width=1\linewidth]{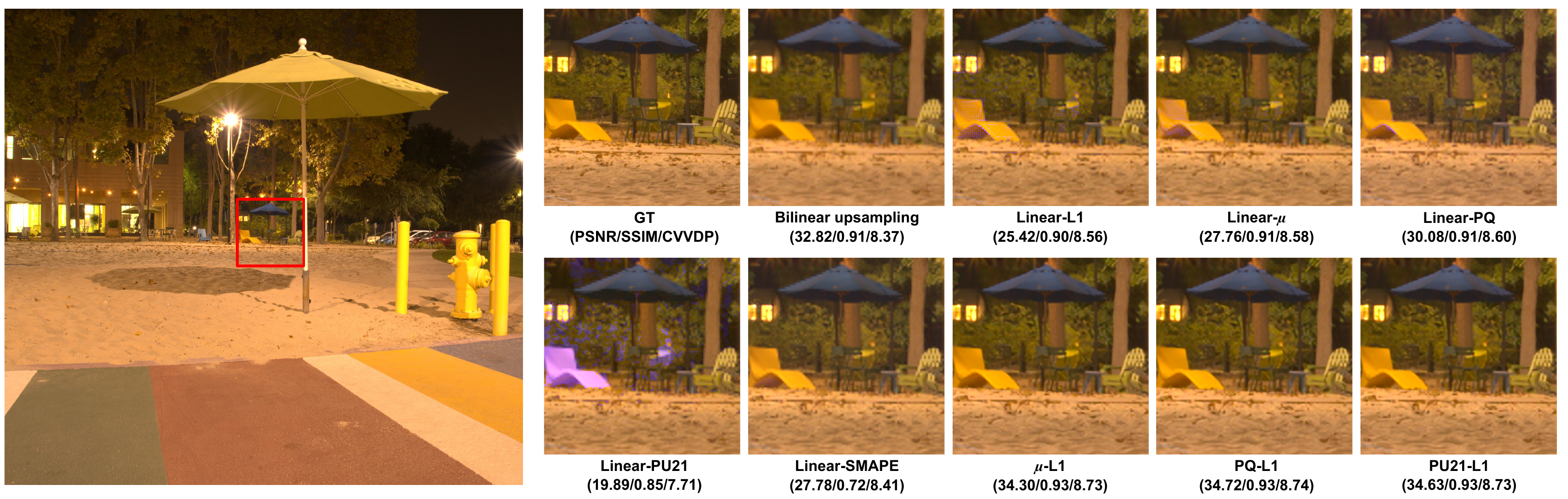} \\
    \includegraphics[width=1\textwidth]{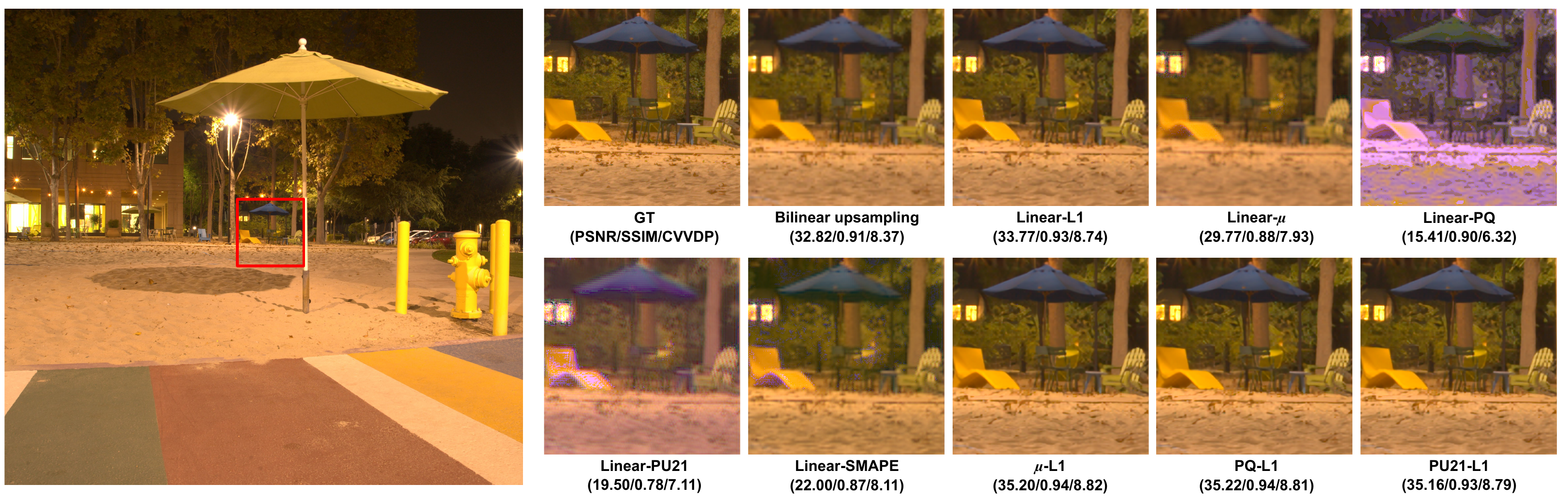}
    \caption{Example results for single-image super-resolution on RAW images with EDSR  \cite{lim2017enhanced} (top) and Real-ESRGAN \cite{wang2021real} (bottom). }
    \label{fig:sisr-vis-res-raw}
\end{figure*}

\begin{figure*}[p]
    \centering
    \includegraphics[width=1\textwidth]{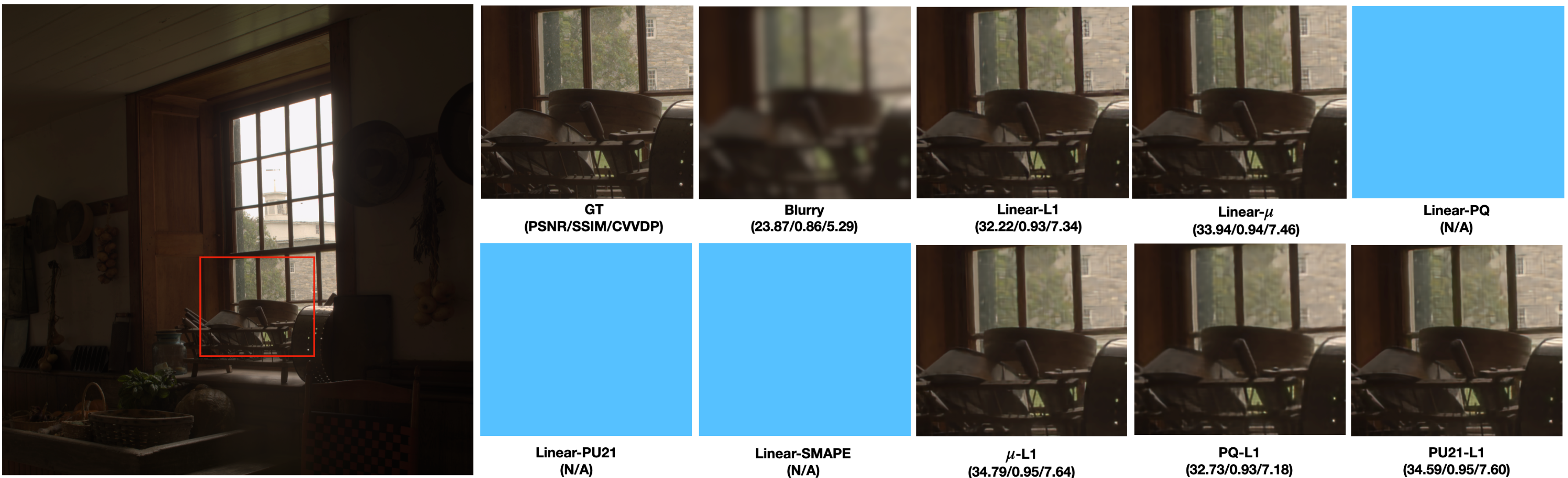}
    \caption{Example results for deblurring with GFNNet. The results are missing for Linear-PQ, Linear-PU21, and Linear-SMAPE, as the networks failed to converge for these configurations.}
    \label{fig:gfnet-vis-res}
\end{figure*}

\begin{figure*}
        \centering
    \includegraphics[width=1\textwidth]{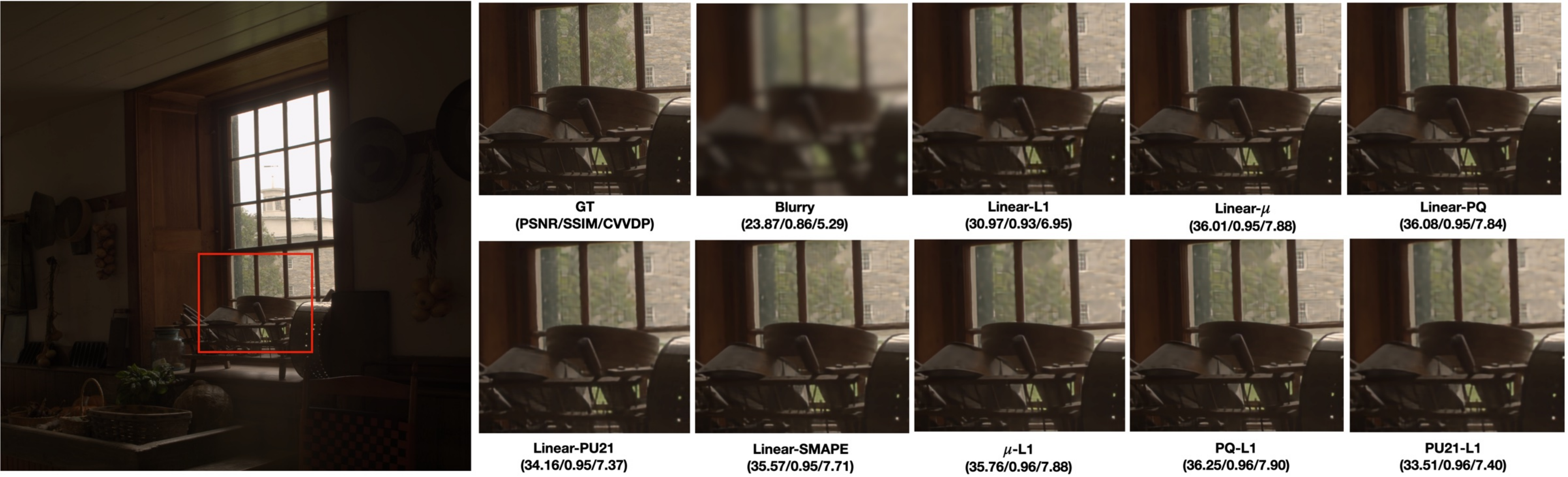}
    \caption{Example results for deblurring with MirNet-v2}
    \label{fig:mirnet-vis-res}
\end{figure*}

\begin{figure*}
        \centering
    \includegraphics[width=1\textwidth]{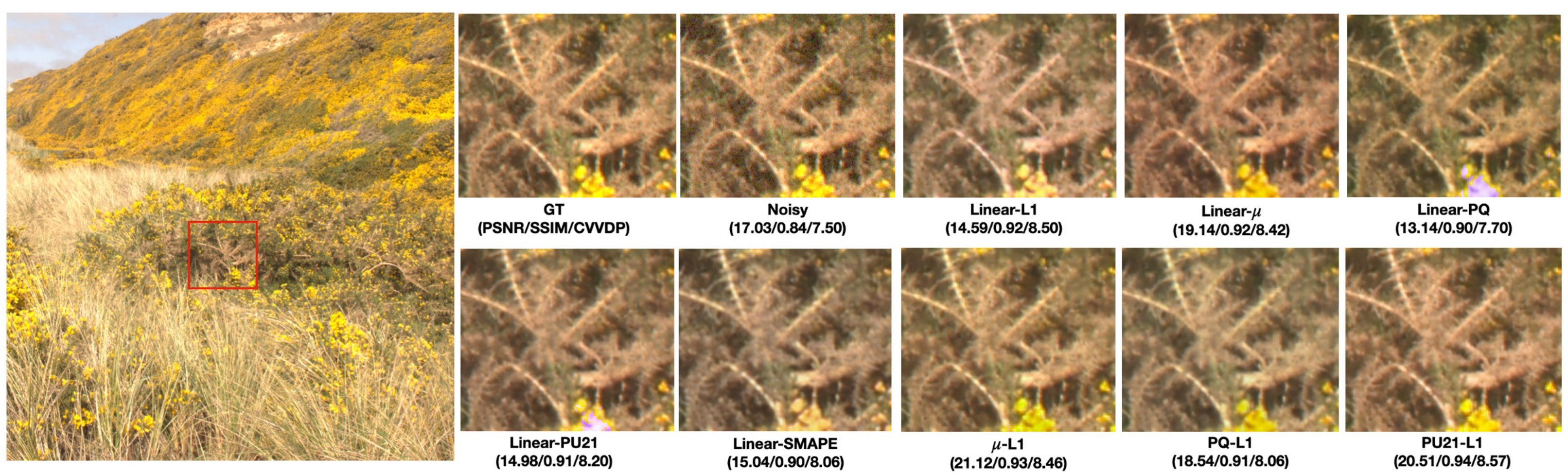} \\
    \includegraphics[width=1\textwidth]{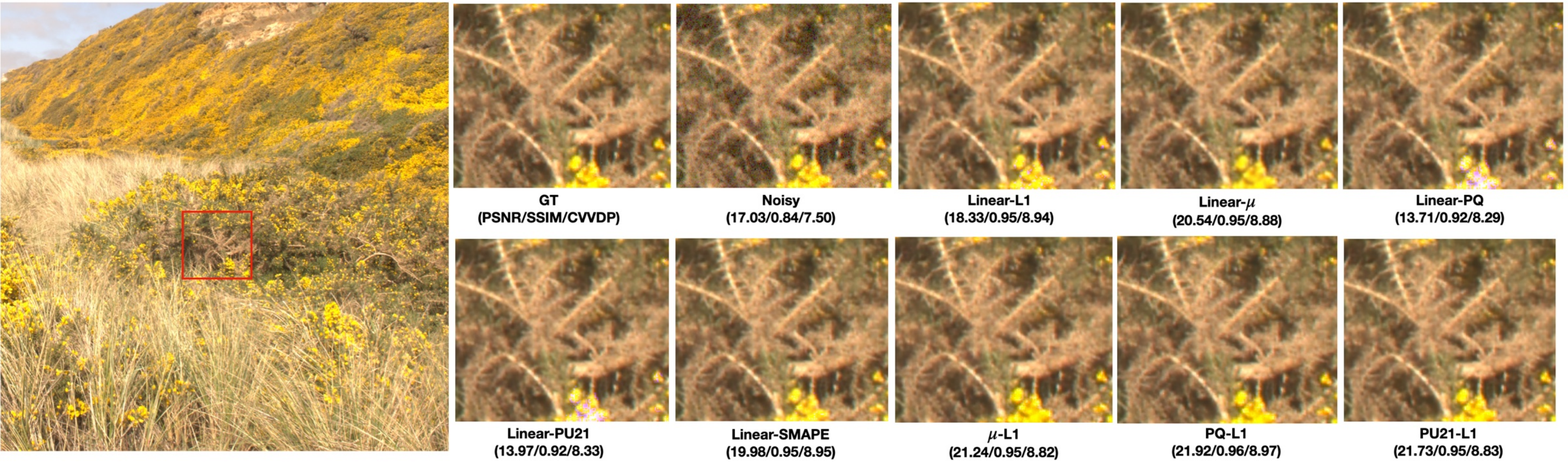}
    \caption{Example results for denoising with DnCNN (top) and SADNet (bottom).}
    \label{fig:dncnn-vis-res}
    \label{fig:sadnet-visres}
    \label{fig:denoising-vis-res}
\end{figure*}

This document contains additional results and details as listed below: 
\begin{itemize}
    \item \tableref{real-ESRGAN-lr} --- learning rates used for Real-ESRGAN
    \item \figref{dataset-histograms} --- histograms over all images and HDR and RAW datasets.
    \item \figref{sisr-vis-res-hdr} --- visual results for single-image super resolution on the HDR image dataset
    \item \figref{sisr-vis-res-raw} --- visual results for single-image super resolution on the RAW image dataset
    \item \figref{gfnet-vis-res} --- visual results for for deblurring with GFNNet
    \item \figref{mirnet-vis-res} --- visual results for for deblurring with MirNet-v2.
    
    \item \figref{denoising-vis-res} --- visual results for denoising --- DnCNN and SADNet
\end{itemize}

\begin{table}[]
    \caption{Learning rates used for Real-ESRGAN.}
    \centering
    {\small
    \begin{tabular}{lc}
    \toprule
         Label & Learning rate\\
         \midrule
         Linear-L1 & $1 \times 10^{-4}$
\\
         PQ-L1 & $1 \times 10^{-4}$\\
         PU21-L1 & $1 \times 10^{-4}$\\
         $\mu$-L1 & $1 \times 10^{-4}$\\
         Linear-PQ & $1 \times 10^{-5}$\\
         Linear-PU21 & $1 \times 10^{-5}$\\
         Linear-$\mu$ & $1 \times 10^{-5}$\\
         Linear-SMAPE & $1 \times 10^{-5}$\\
         \bottomrule
    \end{tabular}
    }
    \label{tab:real-ESRGAN-lr}
    \vspace{-3mm}
\end{table}

%%
% %% The next two lines define the bibliography style to be used, and
% %% the bibliography file.
% \bibliographystyle{ACM-Reference-Format}
% \bibliography{bibliography}

% \end{document}

\end{document}